\documentclass[english,aip,jcp,reprint]{revtex4-2}

\usepackage[T1]{fontenc}
\usepackage[utf8]{inputenc}
\usepackage{babel}
\usepackage{amsmath}
\usepackage{amssymb}
\usepackage{graphicx}
\usepackage[unicode=true,pdfusetitle,
 bookmarks=true,bookmarksnumbered=false,bookmarksopen=false,
 breaklinks=false,pdfborder={0 0 1},backref=false,colorlinks=false]
 {hyperref}

\makeatletter

\newcommand{\lyxmathsym}[1]{\ifmmode\begingroup\def\b@ld{bold}
  \text{\ifx\math@version\b@ld\bfseries\fi#1}\endgroup\else#1\fi}

\usepackage[version=4]{mhchem}
\usepackage{notes2bib}

\bibnotesetup{
note-name = ,
use-sort-key = false
}

\usepackage{cleveref}
\AtBeginDocument{%
\let\ref\cref
}

\crefname{enumi}{issue}{issues}

\@ifundefined{showcaptionsetup}{}{%
 \PassOptionsToPackage{caption=false}{subfig}}
\usepackage{subfig}
\makeatother

\begin{document}
\title{Reproducibility of density functional approximations: how new functionals
should be reported}
\author{Susi Lehtola}
\email{susi.lehtola@alumni.helsinki.fi}

\affiliation{Molecular Sciences Software Institute, Blacksburg, Virginia 24061,
United States}
\affiliation{Department of Chemistry, University of Helsinki, P.O. Box 55, FI-00014
University of Helsinki, Finland}
\author{Miguel A. L. Marques}
\affiliation{Research Center Future Energy Materials and Systems of the University
Alliance Ruhr, Faculty of Mechanical Engineering, Ruhr University
Bochum, Universitätsstraße 150, D-44801 Bochum, Germany}
\begin{abstract}
Density functional theory is the workhorse of chemistry and materials
science, and novel density functional approximations (DFAs) are published
every year. To become available in program packages, the novel DFAs
need to be (re)implemented. However, according to our experience as
developers of Libxc {[}Lehtola et al, SoftwareX 7, 1 (2018){]}, a
constant problem in this task is verification, due to the lack of
reliable reference data. As we discuss in this work, this lack has
lead to several non-equivalent implementations of functionals such
as BP86, PW91, PBE, and B3LYP across various program packages, yielding
different total energies. Through careful verification, we have also
found many issues with incorrect functional forms in recent DFAs.

The goal of this work is to ensure the reproducibility of DFAs: DFAs
must be verifiable in order to prevent reappearances of the abovementioned
errors and incompatibilities. A common framework for verification
and testing is therefore needed. We suggest several ways in which
reference energies can be produced with free and open source software,
either with non-self-consistent calculations with tabulated atomic
densities or via self-consistent calculations with various program
packages. The employed numerical parameters---especially, the quadrature
grid---need to be converged to guarantee a $\lesssim0.1\mu E_{h}$
precision in the total energy, which is nowadays routinely achievable
in fully numerical calculations. Moreover, as such sub-$\mu E_{h}$
level agreement can only be achieved when fully equivalent implementations
of the DFA are used, also the source code of the reference implementation
should be made available in any publication describing a new DFA.
\end{abstract}
\maketitle
\newcommand*\ie{{\em i.e.}}
\newcommand*\eg{{\em e.g.}}
\newcommand*\etal{{\em et al.}}
\newcommand*\citeref[1]{ref. \citenum{#1}}
\newcommand*\citerefs[1]{refs. \citenum{#1}} 
\newcommand*\Citeref[1]{Ref. \citenum{#1}}

\newcommand*\Erkale{{\sc Erkale}}
\newcommand*\Bagel{{\sc Bagel}}
\newcommand*\FHIaims{{\sc FHI-aims}}
\newcommand*\HelFEM{{\sc HelFEM}}
\newcommand*\Gaussian{{\sc Gaussian}}
\newcommand*\LibXC{{\sc LibXC}}
\newcommand*\Orca{{\sc Orca}}
\newcommand*\PySCF{{\sc PySCF}}
\newcommand*\PsiFour{{\sc Psi4}}
\newcommand*\Turbomole{{\sc Turbomole}}

\section{Introduction \label{sec:Introduction}}

Density functional theory\citep{Hohenberg1964_PR_864,Kohn1965_PR_1133}
(DFT) is the workhorse of modern quantum chemistry and materials science.
The key idea in DFT is that the complicated quantum mechanical interactions
of the electrons can be rewritten in terms of the electron density,
only, leading to significantly simpler and more affordable calculations
than those with traditional quantum chemical methods based on direct
solution of the electronic wave function.\citep{Kohn1999_RMP_1253}

As the exact density functional is still unknown, in practice DFT
calculations rely on density functional approximations (DFAs).\citep{Becke2014_JCP_18,Jones2015_RMP_897,Mardirossian2017_MP_2315}
Combined with the development of computer architectures as well as
mathematical algorithms for DFT calculations, modern DFAs have enabled
\emph{ab initio} design of new energy materials\citep{Jain2016_NRM_15004}
and catalysts,\citep{Straker2016_NC_10109} for example. 

Little is known, however, about the form of the exact density functional.
This leaves a great deal of freedom in the construction of DFAs. Therefore,
it is not surprising that a huge variety of DFAs has been proposed
in the literature over the past 60 years: for instance, over 600 DFAs
are available at present in our Libxc library of density functionals.\citep{Lehtola2018_S_1}
Despite the significant number of DFAs already available in the literature,
the development of novel DFAs continues on many fronts. A great many
novel functionals have also appeared in the past three years, as exemplified
by some of the functionals\citep{Verma2019_JPCA_2966,Jana2019_JPCA_6356,Patra2019_PCCP_19639,Furness2020_JPCL_8208,Lehtola2021_JCTC_943,Jana2021_NJP_63007,Jana2021_JCP_24103,Mitrofanov2021_JCP_161103,Ma2022_SA_279,Lebeda2022_PRR_23061}
discussed later in this work.

In order to become available for users of various scientific software
packages, the novel DFAs need to be implemented in those programs.
In the early days of DFT---before the internet and distributed version
control systems became widely available---distributing software was
difficult. As a result, software development typically happened within
a silo mentality: everything (including DFAs) was implemented separately
in each monolithic program package, leading to duplicated effort across
software packages. As a result, disparate choices were made across
various programs, as we will demonstrate in this work for the BP86,\citep{Becke1988_PRA_3098,Perdew1986_PRB_8822}
PW91,\citep{Perdew1991__11,Perdew1992_PRB_6671,Perdew1993_PRB_4978}
PBE,\citep{Perdew1996_PRL_3865,Perdew1997_PRL_1396} and B3LYP\citep{Stephens1994_JPC_11623}
functionals. We also note in this context that \citet{Brakestad2021_JCP_214302}
recently described differences in the BP86 functional between ORCA\citep{Neese2020_JCP_224108}
and MRChem\citep{Wind2023_JCTC_137} of several kcal/mol at the complete
basis set (CBS) limit.

In contrast, in the modern open source paradigm of software development,
common tasks are accomplished via reusable shared modular libraries.\citep{Lehtola2022_WIRCMS_1610}
In the present case of the evaluation of DFAs, the aforementioned
Libxc\citep{Lehtola2018_S_1} is the implementation of choice: Libxc
is used by around 40 electronic structure programs based on various
numerical approaches, such as atomic-orbital basis sets, plane waves,
as well as real-space approaches. Thanks to the modular approach to
software development, new functionals only need to be implemented
in Libxc to become usable in a large number of programs. In addition,
the common library enables access to \emph{exactly the same implementation}
across various numerical approaches (see \citeref{Lehtola2019_IJQC_25968}
for a recent review on numerical approaches), which simplifies, for
instance, the study of numerical precision such as basis set truncation
errors by comparison to fully numerical reference values.

However, in spite of the progress facilitated by the new programming
paradigms, it is essential to verify any new implementation(s) of
DFAs before making them available for use. This verification essentially
boils down to the question of reproducibility: can the main result
of a paper describing a novel DFA---the DFA itself---be exactly
reproduced? 

The ability to verify the DFA is topical thanks to the recent availability
of fully numerical methods that allow the reproduction of reliable
total energies for moderate size systems to guaranteed sub-$\mu E_{h}$
precision.\citep{Lehtola2019_IJQC_25968} Access to such total energies,
determined directly at the CBS limit, enables benchmark studies of
the precision of various numerical approaches.\citep{Jensen2016_PCCP_21145,Jensen2017_JPCL_1449,Lehtola2019_IJQC_25944,Lehtola2019_IJQC_25945,Brakestad2020_JCTC_4874,Lehtola2020_JCP_134108,Brakestad2021_JCP_214302,Lehtola2023_JCTC_2502}
For instance, the basis set truncation error (BSTE) of an approximate
method can be computed as the difference from the CBS limit energy
\begin{equation}
\Delta E=E(\text{approximate})-E(\text{CBS})\ge0,\label{eq:bste}
\end{equation}
affording an unambiguous measure of the precision of the studied numerical
approximation. 

It is clear that \ref{eq:bste} only produces reliable estimates of
the BSTE if the density functional implementations used in the appoximate
and CBS limit calculations agree to very high (sub-$\mu E_{h}$) precision.
As we already mentioned above, this is not always the case, as we
will demonstrate in this work. Therefore, high-precision studies should
always either use the exact same implementation of the density functional
(e.g. using Libxc), or verify that the employed density functional
implementations match to the required precision.

Our general aim with Libxc\citep{Lehtola2018_S_1} is to reproduce
functionals as they were originally employed. The obvious first step
to enable such reproduction of a novel DFA in Libxc and other programs
is to always publish the source code of the reference implementation
as supporting information to the article describing a new DFA. Access
to the source code may be necessary to enable reverse engineering
the implementation that was actually employed, since as we will exemplify
in this work, in many cases the DFA described in an article does not
match the implementation that was actually used to obtain the published
data.

Validation efforts are also greatly aided by unambiguous, reliable
total energies evaluated with the novel functional, as comparing total
energies is easier than comparing implementations in source code.
As we will show in this work by counterexample, matching total energies
to sub-$\mu E_{h}$ precision is sufficient to demonstrate that the
DFAs match, because small changes to the functional parameters often
lead to $\mu E_{h}$ level differences in total energies. However,
we also note here that we have recently shown that many functionals
do not allow facile evaluation of reliable total energies due to numerical
ill-behavedness; we refer the reader to the related literature for
discussion,\citep{Lehtola2022_JCP_174114,Lehtola2023_JPCA_4180,Lehtola2023_JCTC_2502}
and urge functional developers to check and demonstrate that their
new functionals are numerically well-behaved.

The issues with differing implementations of established DFAs in various
program packages is most likely caused by the historical lack of reference
implementations and reliable reference energies. Because these two
issues are still a plague on the implementation and verification efforts
of novel DFAs---as we can attest as longtime developers of Libxc---the
aim of this work is to document various issues we have uncovered in
a number of DFAs, to draw attention to these common issues, and to
prevent them from reoccurring in the future by raising awareness in
the community about the need to be able to verify novel DFAs.

The layout of this work is as follows. Next, in \ref{sec:Theory},
we briefly summarize the mathematical composition of DFAs. Then, in
\ref{sec:problems-with-verification}, we list common problems with
the ways that several DFAs have been reported in the literature. We
suggest three feasible alternative approaches for determining reliable
reference energies in \ref{sec:Feasible-Solutions}: (i) the use of
tabulated wave functions, as well as self-consistent calculations
with (ii) Gaussian basis sets and (iii) fully numerical methods. We
illustrate the usefulness of the three approaches in \ref{sec:Demonstrative-calculations}
by showing how tabulated wave functions can be used to study differences
between density functional implementations, and demonstrating the
need to converge self-consistent calculations to the quadrature grid
limit to allow the determination of reliable reference energies. We
finish with a brief summary and discussion in \ref{sec:Summary-and-Discussion}.
Atomic units are used throughout unless specified otherwise.

\section{Theory \label{sec:Theory}}

In DFT, the total energy is expressed as
\begin{equation}
E[n]=T[n]+V[n]+E_{J}[n]+E_{\text{xc}}[n],\label{eq:Etot}
\end{equation}
where $T$ is the kinetic energy (typically evaluated in terms of
the occupied orbitals as suggested by \citet{Kohn1965_PR_1133}),
$V$ is the nuclear attraction energy, $E_{J}$ is the classical Coulomb
repulsion of the electrons, and $E_{\text{xc}}$ is the quantum mechanical
exchange-correlation energy. Common DFAs express $E_{\text{xc}}$
as

\begin{equation}
E_{\text{xc}}[n]=\int n\epsilon_{\text{xc}}(n_{\uparrow},n_{\downarrow},\nabla n_{\uparrow},\nabla n_{\downarrow},\nabla^{2}n_{\uparrow},\nabla^{2}n_{\downarrow},\tau_{\uparrow},\tau_{\downarrow}){\rm d}^{3}r,\label{eq:Edft}
\end{equation}
where $n_{\uparrow}$ and $n_{\downarrow}$ are the spin-up and spin-down
electron density, and $\tau_{\uparrow}$ and $\tau_{\downarrow}$
are the local kinetic energy densities
\begin{equation}
\tau_{\sigma}=\frac{1}{2}\sum_{i\text{ occupied}}|\nabla\psi_{i\sigma}|^{2}.\label{eq:tau}
\end{equation}
The $\epsilon_{\text{xc}}$ term in \ref{eq:Edft} is the DFA, which
is a (often complicated) mathematical function with known analytical
form. DFAs can be classified on Jacob's ladder based on their ingredients:\citep{Perdew2001_ACP_1}
\begin{itemize}
\item local density approximation (LDA): dependence only on $n_{\uparrow}$
and $n_{\downarrow}$
\item meta-LDA approximation:\citep{Lehtola2021_JCTC_943} dependence on
$n_{\uparrow}$ and $n_{\downarrow}$ as well as $\tau_{\uparrow}$
and $\tau_{\downarrow}$
\item generalized-gradient approximation (GGA): dependence on $n_{\uparrow}$
and $n_{\downarrow}$ as well as their gradients $\nabla n_{\uparrow}$
and $\nabla n_{\downarrow}$
\item meta-GGA approximation: further dependence on the Laplacian $\nabla^{2}n_{\uparrow}$,
$\nabla^{2}n_{\downarrow}$, and/or the local kinetic energy density
$\tau_{\uparrow}$, $\tau_{\downarrow}$
\end{itemize}
Note that two conventions for $\tau_{\sigma}$ exist in the literature:
the one with the physical factor $1/2$, as in our \ref{eq:tau},
and another without it. Several DFAs have been published following
either definition; the actual choice does not matter as long as it
is clear and made consistently.

In addition to a term of the form of \ref{eq:Edft}, some DFAs also
add post-DFT terms to \ref{eq:Etot} such as 
\begin{itemize}
\item exact exchange in either the Hartree--Fock (global hybrids, e.g.
the B3LYP functional\citep{Stephens1994_JPC_11623}) or range-separated
form (range-separated hybrids, e.g. the $\omega$B97X functional\citep{Chai2008_JCP_84106})
\item non-local correlation (e.g. $\omega$B97X-V) or semiempirical dispersion
(e.g. the $\omega$B97X-D3 functional\citep{Lin2013_JCTC_263})
\item post-Hartree--Fock correlation (double hybrids, e.g. the XYG3 functional\citep{Zhang2009_PNAS_4963})
\end{itemize}
Although the need to verify DFAs is transversal to all families, these
additional ingredients will not be discussed further in this work,
because the conclusions of our main analysis also apply to such functionals.
Our main focus is the accurate evaluation of the total energy of \ref{eq:Etot},
and its ramifications on the reproducibility and verification of DFT
calculations.

It is important to note here that the DFA energy of \ref{eq:Edft}
is usually evaluated by quadrature using the scheme pioneered by \citet{Becke1988_JCP_2547};
see our recent work in \citeref{Lehtola2022_JCP_174114} for discussion.
For the present purposes, it suffices to state that the quadrature
is an approximation, which can in principle be made arbitrarily accurate
by using sufficiently many points. It is of utmost importance to study
the convergence of the DFA energy with respect to the size of the
quadrature grid when reporting new DFAs and reference energies for
them, as we will discuss in \ref{subsec:SCF-calculations}.

\section{Common problems with verification \label{sec:problems-with-verification}}

As can be attested by our experience in developing Libxc, verifying
implementations of DFAs is often painstaking, as the usual problems
include the following.
\begin{enumerate}
\item the original article does not report raw total energies, only computed
energy differences such as atomization energies and/or optimized geometries
(for example, refs. \citenum{Neumann1995_CPL_381, Jana2021_NJP_63007, Lebeda2022_PRR_23061})
\label{enu:no-raw-data}
\item the reference values are not fully converged with respect to all numerical
parameters (for example, ref. \citenum{Neumann1995_CPL_381}) \label{enu:not-converged}
\item the reference values are not reported with sufficiently many decimals
(for example, refs. \citenum{Patra2019_PCCP_19639, Jana2021_JCP_24103, Jana2021_NJP_63007, Lebeda2022_PRR_23061})
\label{enu:too-few-decimals}
\item the functional form is incorrect \label{enu:wrong-form}
\begin{enumerate}
\item the article gives the wrong functional form but it is correct in the
reference implementation (for example, refs. \citenum{Proynov1994_PRB_7874, Proynov1998_PRB_12616, Proynov2009_PRA_14103, Proynov2017_PRA_59904, Patra2019_PCCP_19639, mBRbug, Jana2021_NJP_63007, rMGGACbug, Mitrofanov2021_JCP_161103, relPBE0bug})
\bibnotetext[mBRbug]{There's a missing factor of two for the kinetic energy of the homogeneous electron gas in \citeref{Patra2019_PCCP_19639}.}
\bibnotetext[rMGGACbug]{To reproduce the reference implementation of  \citeref{Jana2021_NJP_63007}, the local kinetic energy of \cref{eq:tau} needs to be defined without the factor of one half.}
\bibnotetext[relPBE0bug]{The description of the relPBE0 functional in \citeref{Mitrofanov2021_JCP_161103} omits that the LDA and GGA contributions to correlation need to be scaled by factors of $\approx 0.390518 \dots$ and $\approx 0.60948\dots$, respectively.}
\item the functional form was correct in the paper but not in the reference
implementation (for example, refs. \citenum{Heyd2003_JCP_8207, Heyd2006_JCP_219906, Lebeda2022_PRR_23061, CCaLDAbug})
\bibnotetext[CCaLDAbug]{To reproduce the reference implementation of  \citeref{Lebeda2022_PRR_23061}, an additional factor of $2^{2/3}$ needs to be added in the definition of $\alpha$.}
\end{enumerate}
\item too few details are given on how the data was actually obtained (for
example, refs. \citenum{Neumann1995_CPL_381, Perdew1999_PRL_2544, Seidl2000_PRL_5070, Haas2011_PRB_205117, Jana2018_CPL_1, Jana2018_PCCP_8999, Jana2019_JPCA_6356})
\label{enu:too-few-details}
\item the parameter values in the implementation are different from the
paper (for example, \citeref{Perdew1996_PRL_3865, PBEbug, Adamo1998_JCP_664, mPW91bug, Boese2000_JCP_1670, hcthbug, Hoe2001_CPL_319, O3LYPbug, Peverati2012_JCTC_2310, N12bug, Peverati2012_PCCP_16187, N12SXbug, Ma2022_SA_279, GAS22bug})
\label{enu:different-parameters}

\bibnotetext[PBEbug]{The implementation of \citeref{Perdew1996_PRL_3865} employs a more precise value for $\mu$ than the one given in the paper.}
\bibnotetext[mPW91bug]{In commit 4b9609d1c57 of the source code of the NWChem implementation, dated 19 Feb 2003, Edoardo Apra comments that Adamo has confirmed that there is a typo in the JCP paper; $b = 0.00426$ instead of 0.0046 given in the text of \citeref{Adamo1998_JCP_664}, also the exponent is 3.72 and not 3.73 as given in the manuscript. See \url{https://github.com/nwchemgit/nwchem/blob/master/src/nwdft/xc/xc\_xmpw91.F} (accessed 1 June 2022).}
\bibnotetext[hcthbug]{The $c_5$ parameter of HCTH/147 was given with the wrong sign in \citet{Boese2000_JCP_1670}. Another set of parameter values are given in \citet{Boese2003_JCP_3005} that agree with the ones in \citeref{Boese2000_JCP_1670} for $c_1$--$c_3$ and $c_9$--$c_{15}$ but have small differences for the $c_4$--$c_8$ coefficients. \Citeref{Boese2003_JCP_3005} gives (in addition to the correct sign for $c_5$) one more decimal for $c_4$--$c_7$ than \citeref{Boese2000_JCP_1670} but also a differently rounded value for $c_8$. The original implementation in CADPAC appears to use still more decimals.}
\bibnotetext[O3LYPbug]{The parameters of \citeref{Hoe2001_CPL_319} do not reproduce the data of the paper; see \url{http://www.ccl.net/chemistry/resources/messages/2008/10/09.007-dir/index.html} (accessed 1 June 2022).}
\bibnotetext[N12bug]{The same-spin and opposite-spin correlation coefficients are interchanged in \citeref{Peverati2012_JCTC_2310}.}
\bibnotetext[N12SXbug]{The same-spin and opposite-spin correlation coefficients are interchanged in \citeref{Peverati2012_PCCP_16187}. Moreover, the exchange functional coefficients need to be transposed.}
\bibnotetext[GAS22bug]{The Jupyter notebook code of \citeref{Ma2022_SA_279} employs many more decimals for the parameters than given in the paper.}
\end{enumerate}
We will now proceed to explain why these are problems for verification
of DFAs.

The problem with \ref{enu:no-raw-data} is that energy differences
or optimized geometries tend to exhibit systematic error cancellations.
They are therefore less sensitive to the functional form and to the
values of the employed parameters than the total energy is. Although
physical properties and chemistry often also benefit from such error
cancellation,\bibnote{This error cancellation is also the reason why slightly different implementations of a DFA still tend to yield similar geometries and excitation energies, for example.}
one needs to compare raw total energies to ensure that two DFA implementations
are equivalent.

Similarly for \ref{enu:not-converged}, the typical problem is that
a default quadrature grid has been used to evaluate the density functional.
This almost always means that the data is not converged to the precision
necessary to compare two different implementations, as the default
grids tend to be chosen to maximize speed while maintaining a sufficient
level of precision for chemical applications. However, default grids
don't always guarantee such a level of precision even for standard
calculations.\citep{Martin2001_CPC_189,Papas2006_JMST_175,Wheeler2010_JCTC_395}

As was already remarked above in \ref{sec:Theory}, the quadrature
error needs to be made insignificant for proper comparisons to take
place. As we have discussed in \citeref{Lehtola2022_JCP_174114} for
fixed densities, and as will be demonstrated with self-consistent
calculations in \ref{subsec:SCF-calculations}, this can require hundreds
of radial grid points in the case of many functionals; polyatomic
calculations will also require the use of large angular grids to yield
fully converged total energies.

\Cref{enu:too-few-decimals} is that many works report total energies
in Hartrees with only three-decimal precision. Such 1 $\text{m}E_{h}$
precision corresponds to roughly $30$ meV, which is large enough
in our experience to hide errors both in the functional form (\ref{enu:wrong-form})
as well as discrepancies in parameter values (\ref{enu:different-parameters}).
The best practice is to report reference total energies to 1 $\mu E_{h}$
precision or better.

\Cref{enu:wrong-form} is self-explanatory: the functional is not
what was published. In the present authors' opinion, the correct implementation
should reproduce the results in the paper, unless the original authors'
implementation was incorrect and the results in the paper have been
rectified with a further erratum. A famous example of this is the
Heyd--Scuseria--Ernzerhof (HSE) functional,\citep{Heyd2003_JCP_8207,Heyd2006_JCP_219906}
whose HSE03 variant corresponds to the original erroneous implementation,
and the HSE06 variant to the rectified implementation, which is the
functional that was supposed to have been used in the original paper. 

The HSE functionals are actually infamous for reproducibility: various
codes have implemented the functionals in dissimilar manners. Specifically,
there is an enormous mess concerning the values of the range-separation
parameter $\omega$ in the HSE functionals in the literature, as well
as in the available implementations. To rehash, the original paper\citep{Heyd2003_JCP_8207}
stated that the range-separation parameter $\omega^{\text{HF}}=0.15=\omega^{\text{PBE}}$
was used for both the HF and the PBE parts of the functional. However,
due to an error in the code, the real value used was $\omega^{\text{HF}}=0.15/\sqrt{2}\approx0.1061$
and $\omega^{\text{PBE}}=0.15\sqrt[3]{2}\approx0.1890$, according
to the erratum published a few years later.\citep{Heyd2006_JCP_219906} 

\citet{Krukau2006_JCP_224106} tried to clarify the situation, and
called the original choice of parameters with $\omega^{\text{HF}}\neq\omega^{\text{PBE}}$
HSE03, and the functional where $\omega^{\text{HF}}=\omega^{\text{PBE}}$
HSE06. By testing several properties for atoms, molecules, and solids,
\citet{Krukau2006_JCP_224106} determined the best value $\omega^{\text{HF}}=\omega^{\text{PBE}}=0.11$
for the HSE06 form, differring from the values described in \citerefs{Heyd2003_JCP_8207}
and \citenum{Heyd2006_JCP_219906}.

Now, HSE06 in Quantum Espresso\citep{Giannozzi2020_JCP_154105} employs
the value $\omega^{\text{HF}}=\omega^{\text{PBE}}=0.106$, which is
clearly not the $\omega^{\text{HF}}=\omega^{\text{PBE}}=0.15$ of
the HSE06 of \citet{Heyd2006_JCP_219906}, nor the reoptimized value
$\omega^{\text{HF}}=\omega^{\text{PBE}}=0.11$ of \citet{Krukau2006_JCP_224106}. 

HSE06 in VASP,\citep{Hafner2008_JCC_2044} in turn, employs $\ensuremath{\omega^{\text{HF}}=\omega^{\text{PBE}}=0.2\lyxmathsym{\AA}^{-1}\ensuremath{\approx}0.1058}$,
which is similar to (but not the same as!) the value used in Quantum
Espresso, and also disagrees with the values used by \citet{Heyd2006_JCP_219906}
and \citet{Krukau2006_JCP_224106}. Even more surprisingly, HSE03
in VASP employs $\omega^{\text{HF}}=\omega^{\text{PBE}}=0.3\lyxmathsym{\AA}^{-1}\ensuremath{\approx}0.1587$,
breaking with the terminology of HSE03 \emph{vs.} HSE06 suggested
by \citet{Krukau2006_JCP_224106} where HSE03 is defined by $\omega^{\text{HF}}\neq\omega^{\text{PBE}}$.

The situation with the HSE functionals is complicated even further
by the introduction of scaling functions\citep{Heyd2004_JCP_7274,Henderson2008_JCP_194105,Henderson2009_JCP_44108}
to force the functional to obey the local version of the Lieb--Oxford\citep{Lieb1981_IJQC_427}
bound. In fact, we find several modern implementations using different
scaling functions, which lead to slightly different results.

Because of the above discrepancies, it is unfortunately the end user's
responsibility to check whether the implementations in any two codes
are sufficiently similar to enable meaningful comparison of the results.
The standard modular implementations provided by Libxc\citep{Lehtola2018_S_1}
are thereby invaluable, as they enable apples-to-apples comparisons
of these types of functionals, as well.

In \ref{enu:too-few-details}, sufficient details have not been provided
on how the data reported in the paper were obtained. In practice,
this may mean that the basis set and/or the quadrature grid that were
used for the calculations has not been specified. Without this information,
the results cannot be reproduced and their quality cannot be judged,
making the data worthless.

\Cref{enu:different-parameters} is present in many older functionals,
where small differences in the parameter values that usually arise
from truncation cause differences typically of the order $10^{-4}E_{h}$,
which are thereby noticeable when the calculations are tightly converged;
see \ref{sec:Demonstrative-calculations} for examples.

\section{Simple Solutions \label{sec:Feasible-Solutions}}

\subsection{Tabulated wave functions \label{subsec:Tabulated-wave-functions}}

If the DFA has the form of \ref{eq:Edft}, it may suffice to report
non-self-consistent energies computed on top of atomic Hartree--Fock
wave functions. Our reasoning is the following: when combined with
symbolic algebra as in Libxc\citep{Lehtola2018_S_1} or automatic
differentiation as in XCFun\citep{Ekstroem2010_JCTC_1971}, the correctness
of the energy evaluation---which is ensured by the non-self-consistent
single-point evaluation---already ensures that the gradient of the
energy is also correct, since computer algebra systems and automatic
differentiation toolkits are not expected to give faulty derivatives.
We also note here that if a functional's energy gradient has not been
implemented correctly, the result of a self-consistent calculation
is a total energy that is higher than the true ground state energy,
provided that the calculation is variational, like modern Gaussian-basis
and fully numerical calculations are.\citep{Lehtola2019_IJQC_25968,Lehtola2020_M_1218}

Although tabulated Hartree--Fock wave functions are often thought
to be synonymous with those of \citet{Clementi1974_ADNDT_177}, which
still appear to be used by functional developers, we note that more
accurate atomic Hartree--Fock wave functions have been reported in
the literature.\citep{Koga1999_IJQC_491,Koga2000_TCA_411} We made
these wave functions accessible in a simple and easy-to-use Python
package called AtomicOrbitals\citep{Furness__} in \citeref{Lehtola2022_JCP_174114}.
This package is interfaced with Libxc, and it also allows for easy
access to atomic densities and quadrature grids, thus allowing the
use of custom implementations of novel DFAs, as well. We recently
employed the wave functions of \citet{Koga1999_IJQC_491} to study
the numerical well-behavedness of DFAs in \citeref{Lehtola2022_JCP_174114},
and found many recent DFAs to be ill-behaved.

\subsection{Self-consistent calculations \label{subsec:Self-consistent-calculations}}

Non-self-consistent calculations at fixed reference densities likely
suffice to determine whether two DFA implementations agree. However,
non-self-consistent calculations do not afford a full examination
of the numerical stability of a DFA: in a self-consistent field (SCF)
calculation, the electron density can adapt to features of the DFA,
and a poorly behaved density functional can exhibit numerical pathologies
in SCF calculations in extended basis sets.\citep{Lehtola2022_JCP_174114,Lehtola2023_JCTC_2502,Lehtola2023_JPCA_4180}

A good test system for studying various kinds of DFAs should have
a well-behaved electron configuration. The test system should not
have low-lying excited states onto which the SCF procedure could converge,
as such saddle-point convergence would unnecessarily complicate the
determination of whether a reimplementation of a DFA is faithful to
the original reference implementation. For similar reasons, the test
system should also not exhibit spatial or spin symmetry breaking.
The ideal systems therefore have either half-closed or fully closed
electronic shells. Moreover, since errors in the energy are likely
extensive, the best choice is to focus on the smallest possible bound
systems: atoms. 

Studying atoms comes with significant added benefits for the density
functional comparison. Typical basis set expansions are much better
behaved in the case of single atoms, obviating the need for approaches
to choose an unambiguous basis for carrying out the electronic structure
calculation,\citep{Lehtola2019_JCP_241102} and allowing the use of
extended benchmark-quality basis sets,\citep{Lehtola2020_JCP_134108}
for instance. In addition, calculations on atoms can also easily be
carried out with fully numerical methods: the high amount of symmetry
inherent in atomic problems enabled accurate numerical calculations
already over 60 years ago.\citep{Lehtola2019_IJQC_25968}

The N and Ne atoms are excellent choices for verification purposes.
They are light atoms and have half-closed $1s^{2}2s^{2}2p^{3}$ (quartet)
and closed-shell $1s^{2}2s^{2}2p^{6}$ (singlet) configurations, respectively,
while being sufficiently heavy to exhibit significant electron correlation.
Moreover, these atoms are sufficiently well-behaved to not cause numerical
issues for most functionals, unlike the lighter atoms with half-closed
shells: hydrogen only has a single electron, while lithium has a pronouncedly
diffuse electron distribution which is problematic for many functionals.\citep{Lehtola2022_JCP_174114}

We note that it is important to include both spin-restricted (neon)
and spin-unrestricted (nitrogen) systems in the verification, because
the verification is usually easiest to start with the former, as the
latter type of systems tend to be more complicated due to the spin
polarization.

For similar reasons, two sets of reference energies should be included
for new exchange-correlation functionals: self-consistent energies
for the exchange-only approximation, as well as for calculations including
both exchange and correlation. The exchange functional is energetically
much more important than the correlation functional, and the correctness
of the correlation component should only be checked once the correctness
of the usually much simpler exchange part has been certified. 

We will now proceed to discuss two types of numerical approaches for
carrying out self-consistent calculations on these atoms, affording
well-defined reference energies to high precision.

\subsubsection{Gaussian-basis calculations \label{subsec:Gaussian-basis-calculations}}

Gaussian-basis calculations offer an easy choice for self-consistent
calculations. Gaussian basis sets of various sizes are available,\citep{Jensen2013_WIRCMS_273,Hill2013_IJQC_21}
ranging from minimal basis sets\citep{Hehre1969_JCP_2657} to extended
basis sets designed especially for benchmark quality calculations.\citep{Lehtola2020_JCP_134108}
A large number of Gaussian-basis programs are likewise available for
performing the necessary calculations. In addition to established
commercial packages, several programs that are free and open source
software (FOSS) have also become available in recent years.\citep{Lehtola2022_WIRCMS_1610}
Here we especially want to mention PySCF\citep{Sun2020_JCP_24109}
and Psi4,\citep{Smith2020_JCP_184108} which are both interfaced to
Libxc and enable efficient density functional calculations.

The actual basis set used to perform the calculation is not as important
as ensuring that the basis set is unambiguously defined. For instance,
the Dunning cc-pVXZ basis sets\citep{Dunning1989_JCP_1007} have famous
discrepancies across program packages that mainly arise from different
ways to compute two-electron integrals. The cc-pVXZ basis sets are
generally contracted,\citep{Raffenetti1973_JCP_4452} and many programs
designed for segmented contractions employ modified variants of these
basis sets for improved computational efficiency.\citep{Hashimoto1995_CPL_190,Davidson1996_CPL_514}
There are also discrepancies between versions of the basis sets included
in various programs, as exemplified by the recent work of \citet{Kermani2023_CPL_140575}.

For this reason, we recommend employing basis sets downloaded from
the Basis Set Exchange,\citep{Pritchard2019_JCIM_4814} and enclosing
the used basis set in the Supporting Information. The Hartree--Fock
total energy should also be reported, as it can be used for an independent
test of whether the basis set really is the same in the calculations
in various programs.

\subsubsection{Fully numerical calculations \label{subsec:Fully-numerical-calculations}}

Fully numerical calculations\citep{Lehtola2019_IJQC_25968} go one
step further from Gaussian-basis calculations: a flexible numerical
basis set allows converging total energies to sub-$\mu E_{h}$ precision
from the CBS for the given density functional. For example, the freely
available open source \HelFEM{} program\citep{Lehtola2019_IJQC_25945,Lehtola2019_IJQC_25944,Lehtola2020_PRA_12516,Lehtola2023_JCTC_2502,Lehtola2023_JPCA_4180}
employs the finite element method, which affords a quick approach
to the CBS limit. \HelFEM{} is interfaced to Libxc and supports LDA,
GGA and meta-GGA calculations on atoms and diatomic molecules including
hybrid functionals. In the case of atoms, also range-separated functionals
are supported.\citep{Lehtola2020_PRA_12516} Fully numerical Hartree--Fock
calculations are likewise possible in \HelFEM{} within the single-determinant
or fractional-occupation approach. Fully numerical calculations on
atoms with \HelFEM{} have been extensively discussed in \citerefs{Lehtola2019_IJQC_25945},
\citenum{Lehtola2023_JPCA_4180}, \citenum{Lehtola2023_JCTC_2502},
and \citenum{Lehtola2020_PRA_12516}, to which we refer to further
details.

\section{Demonstrative calculations \label{sec:Demonstrative-calculations}}

\subsection{Studies at fixed density \label{subsec:Studies-at-fixed}}

In this subsection, we will demonstrate the effects that small changes
to the parameters or to the functional form of the PBE, P86, PW91,
PW92, and B3LYP functionals have on the resulting total energy, employing
the tabulated wave functions discussed in \ref{subsec:Tabulated-wave-functions}.
The total energies are evaluated with the scheme discussed in \citeref{Lehtola2022_JCP_174114}
employing the default $N=2000$ radial quadrature points.

\begin{table}
\begin{tabular}{lrr}Functional & $ E_x(\text{N})/E_h $ & $ E_x(\text{Ne})/E_h $\\
\hline
\hline
\texttt{gga\_x\_pbe} & $ -3.9560257 $ & $ -4.4483087 $\\
\texttt{gga\_x\_pbe\_mod} & $ -3.9560281 $ & $ -4.4483114 $\\
\texttt{gga\_x\_pbe\_gaussian} & $ -3.9560181 $ & $ -4.4483003 $\\
\end{tabular}
\caption{Exchange energies of the N and Ne atoms for variants of the PBE exchange functional, employing tabulated Hartree--Fock wave functions\cite{{Koga1999_IJQC_491}} evaluated with AtomicOrbitals\cite{{Furness__}} and a 2000 point radial quadrature with the default scheme of \citeref{Lehtola2022_JCP_174114}. Libxc keywords are used to identify the functionals.}
\label{tab:pbe-x}
\end{table}

\begin{table}
\begin{tabular}{lrr}Functional & $ E_c(\text{N})/E_h $ & $ E_c(\text{Ne})/E_h $\\
\hline
\hline
\texttt{gga\_c\_p86} & $ -0.1490949 $ & $ -0.1762603 $\\
\texttt{gga\_c\_p86\_ft} & $ -0.1491707 $ & $ -0.1763458 $\\
\texttt{gga\_c\_p86vwn} & $ -0.1589501 $ & $ -0.1782665 $\\
\texttt{gga\_c\_p86vwn\_ft} & $ -0.1590260 $ & $ -0.1783520 $\\
\end{tabular}
\caption{Correlation energies of the N and Ne atoms for variants of the P86 correlation functional, employing tabulated Hartree--Fock wave functions\cite{{Koga1999_IJQC_491}} evaluated with AtomicOrbitals\cite{{Furness__}} and a 2000 point radial quadrature with the default scheme of \citeref{Lehtola2022_JCP_174114}. Libxc keywords are used to identify the functionals.}
\label{tab:p86-c}
\end{table}

\paragraph{PBE}

Exchange functionals are typically written in terms of enhancement
functions
\begin{equation}
E_{x}=\sum_{\sigma}\int\epsilon_{x}^{\text{LDA}}(n_{\sigma})F_{x}(s_{\sigma}){\rm d}^{3}r,\label{eq:enhancement}
\end{equation}
where
\begin{equation}
s_{\sigma}=\frac{x_{\sigma}}{2(6\pi^{2})^{1/3}}\label{eq:s-sigma}
\end{equation}
is the standard expression for the reduced gradient where
\begin{equation}
x_{\sigma}=|\nabla n_{\sigma}|/n_{\sigma}^{4/3}\label{eq:x-sigma}
\end{equation}
is a ``bare'' reduced gradient which is also used in some density
functionals.

The PBE exchange functional\citep{Perdew1996_PRL_3865,Perdew1997_PRL_1396}
is defined by the simple enhancement factor
\begin{align}
F_{\text{x}}^{\text{PBE}}(s_{\sigma}) & =1+\kappa-\frac{\kappa}{1+\frac{\mu s_{\sigma}^{2}}{\kappa}}=1+\kappa\left(1-\frac{\kappa}{\kappa+\mu s_{\sigma}^{2}}\right)\label{eq:PBEx}
\end{align}
that depends only on two parameters: $\kappa$ and $\mu$, which control
the $s_{\sigma}\to\infty$ asymptotic value of the enhancement function
and the coefficient of the $s_{\sigma}^{2}$ term in the $s_{\sigma}\to0$
limit, respectively.

However, at least two variants of the PBE exchange functional can
be found in actual implementations. Although the parameter $\kappa$
typically has the value $\kappa=0.804$, there are differences in
the value of $\mu$. The problem is that there are several definitions
in the paper of \citet{Perdew1996_PRL_3865}: first, $\mu=\beta\pi^{2}/3$
with $\beta=0.066725$; second, $\mu=0.21951$ (the first choice would
give $\mu=0.21952$); and third, the value $\beta=0.06672455060314922$
used in Burke's reference implementation. Libxc\citep{Lehtola2018_S_1}
follows the reference implementation and employs the precise value
in \texttt{gga\_x\_pbe}. In contrast, XCFun\citep{Ekstroem2010_JCTC_1971}
employs the first option; this variant is available in Libxc as \texttt{gga\_x\_pbe\_mod}.

We also comment on a third implementation: for historical reasons,
the implementation of PBE exchange in the \Gaussian{} program is
equivalent to the choice $\kappa=0.804000423825475$ and $\mu=0.219510240580611$.\bibnote{Giovanni Scalmani, private communication (2022).}
This variant is also available in Libxc as \texttt{gga\_x\_pbe\_gaussian.}

These different choices for $\kappa$ and $\mu$ show up as detectable
differences in the total energy, as is demonstrated by the data in\textbf{
}\ref{tab:pbe-x}. Even though the value of $\beta$ differs by just
7 parts-per-million between \texttt{gga\_x\_pbe }and\texttt{ gga\_x\_pbe\_mod},
the resulting differences in total energies are still significant.

We note here that most functionals in Libxc allow overriding the parameter
values with \texttt{the xc\_func\_set\_ext\_params() }function, allowing
the use of PBE exchange with arbitrary values for $\kappa$ and $\mu$,
for example.

\paragraph{P86}

A similar issue exists in Perdew's 1986 correlation functional (P86),\citep{Perdew1986_PRB_8822,Perdew1986_PRB_7406}
which together with Becke's 1988 exchange functional\citep{Becke1988_PRA_3098}
(B88) forms the famous BP86 exchange-correlation functional. Although
P86 relies on several fitted parameters, it also depends on a numerical
constant $1.745\tilde{f}$ with $\tilde{f}=0.11$ given in the paper. 

However, it turns out that the numerical factor $1.745$ is in fact
an approximate value for 
\begin{equation}
(9\pi)^{1/6}\eqsim1.74541506\dots\label{eq:lm-9pi}
\end{equation}
 that originates from the Langreth--Mehl correlation functional,\citep{Langreth1981_PRL_446,Hu1985_PS_391}
which was the basis for P86. Some implementations of P86 opt to use
the exact value of \ref{eq:lm-9pi}. 

Both implementations are available in Libxc; the default version,
\texttt{gga\_c\_p86}, employs the value 1.745 specified in the paper,
while \texttt{gga\_c\_p86\_ft} employs $(9\pi)^{1/6}$. This $\mathcal{O}(10^{-4})$
difference in the value of the numerical constant is again visible
in total energies, if they are reported to sufficient precision, as
is seen from \ref{tab:p86-c}.

Note that the P86 functional is based on the PZ LDA correlation functional,
which was found to be numerically ill-behaved in \citeref{Lehtola2022_JCP_174114}
due to its poor convergence to the quadrature limit.\citep{Lehtola2022_JCP_174114}
This also means that the energies given in \ref{tab:p86-c} for the
\texttt{gga\_c\_p86} and \texttt{gga\_c\_p86\_ft} functionals are
unlikely to be converged to sub-$\mu E_{h}$ precision.\citep{Lehtola2022_JCP_174114}
In contrast, the variants of P86 based on the VWN functional were
found to be numerically well-behaved in \citeref{Lehtola2022_JCP_174114}.
The data for these \texttt{gga\_c\_p86vwn} and \texttt{gga\_c\_p86vwn\_ft}
functionals in \ref{tab:p86-c} likewise illustrate the effect of
the truncation of $(9\pi)^{1/6}$. Note that the VWN and PZ functionals
employ different spin interpolation formulas, leading to larger differences
in energy for the spin-polarized nitrogen atom than for the spin-restricted
neon atom.

\paragraph{PW91}

The 1991 Perdew--Wang exchange functional\citep{Perdew1991__11,Perdew1992_PRB_6671,Perdew1993_PRB_4978}
(PW91) is another interesting case. The enhancement factor for this
functional as described in \citerefs{Perdew1991__11} and \citenum{Perdew1992_PRB_6671}
reads as \begin{widetext}
\begin{equation}
F(s_{\sigma})=\frac{1+0.19645s_{\sigma}\sinh^{-1}(7.7956s_{\sigma})+(0.2743-0.1508e^{-100s_{\sigma}^{2}})s_{\sigma}^{2}}{1+0.19645s_{\sigma}\sinh^{-1}(7.7956s_{\sigma})+0.004s_{\sigma}^{4}}.\label{eq:PW91}
\end{equation}
\end{widetext} This functional, which is available in Libxc as \texttt{gga\_x\_pw91},
appears to contain 6 numerical parameters. However, it turns out that
this functional can also be written in a different form: according
to the literature,\citep{Gill1996_MP_433,Adamo1998_JCP_664} the enhancement
factor of PW91 exchange can also be written as
\begin{equation}
F(x_{\sigma})=1+|A_{x}|^{-1}\frac{bx_{\sigma}^{2}-(b-\beta)x_{\sigma}^{2}\exp(-cx_{\sigma}^{2})-10^{-6}x_{\sigma}^{d}}{1+6bx_{\sigma}\sinh^{-1}x_{\sigma}+\frac{10^{-6}x_{\sigma}^{d}}{|A_{x}|}}\label{eq:PW91-2nd}
\end{equation}
where $\beta=5(36\pi)^{-5/3}$, $b=0.0042$, $c=1.6455$, $d=4$,
and
\begin{equation}
A_{x}=-\frac{3}{2}\left(\frac{3}{4\pi}\right)^{1/3}.\label{eq:Ax}
\end{equation}
This form apparently only has three numerical parameters: $b$, $c$,
and $d$, since $\beta$ and $A_{x}$ are constants ($d$ could also
be thought to be fixed by the choice of the functional form). Some
programs implement the PW91 exchange functional using the form of
\ref{eq:PW91-2nd}, and this form is available in Libxc as \texttt{gga\_x\_pw91\_mod}.

However, \ref{eq:PW91-2nd,eq:PW91} are not fully equivalent. Rewriting
\ref{eq:PW91-2nd} in the form of \ref{eq:PW91} by collecting the
terms over the same denominator and using \ref{eq:s-sigma} to convert
the dependence on $x_{\sigma}$ to that on $s_{\sigma}$, one finds
that the above choices for the parameters for \ref{eq:PW91-2nd} instead
lead to the approximate form\begin{widetext}
\begin{equation}
F(s_{\sigma})\approx\frac{1+0.196447965s_{\sigma}\sinh^{-1}(7.795554179s_{\sigma})+(0.274293104-0.150836314e^{-99.998129199s_{\sigma}^{2}})s_{\sigma}^{2}}{1+0.196447965s_{\sigma}\sinh^{-1}(7.795554179s_{\sigma})+0.003968803s_{\sigma}^{4}}.\label{eq:PW91-2nd-rewrite}
\end{equation}
\end{widetext} \Cref{eq:PW91-2nd-rewrite} is clearly different from
\ref{eq:PW91} and will therefore obviously give a different numerical
result, as is also clear from the results in \ref{tab:pw91-x}. The
coefficient of the $s_{\sigma}^{4}$ term differs in the two implementations
by a whopping 0.8 \%, the other parameters exhibiting differences
that are smaller by 1--3 orders of magnitude.

We note here that the book in which \citeref{Perdew1991__11} was
published is not online and has been out of print for decades, and
the present authors failed to access it despite prolonged efforts.
Because of this and other similar issues, our recommendation is to
publish novel functionals as journal articles that are more likely
to remain accessible in the future.\bibnote{We have finally received a copy of \citeref{Perdew1991__11} from colleagues abroad, because of an analogous disclaimer in a preprint of this work.}

\begin{table}
\begin{tabular}{lrr}Functional & $ E_x(\text{N})/E_h $ & $ E_x(\text{Ne})/E_h $\\
\hline
\hline
\texttt{gga\_x\_pw91} & $ -3.9561253 $ & $ -4.4528458 $\\
\texttt{gga\_x\_pw91\_mod} & $ -3.9565874 $ & $ -4.4532707 $\\
\end{tabular}
\caption{Exchange energies of the N and Ne atoms for variants of the PW91 exchange functional, employing tabulated Hartree--Fock wave functions\cite{{Koga1999_IJQC_491}} evaluated with AtomicOrbitals\cite{{Furness__}} and a 2000 point radial quadrature with the default scheme of \citeref{Lehtola2022_JCP_174114}. Libxc keywords are used to identify the functionals.}
\label{tab:pw91-x}
\end{table}

\paragraph{PW92}

The 1992 Perdew--Wang LDA correlation functional\citep{Perdew1992_PRB_13244}
(PW92) is another interesting case. The PW92 functional employs the
spin interpolation formula of \citet{Vosko1980_CJP_1200} with
\begin{equation}
f(\zeta)=\frac{[(1+\zeta)^{4/3}+(1-\zeta)^{4/3}-2]}{2^{4/3}-2}.\label{eq:pw92-fzeta}
\end{equation}
Even though the exact value of $f''(0)$, which is used in the functionals
of \citet{Vosko1980_CJP_1200}, is easy to evaluate to
\begin{equation}
f''(0)=\frac{4}{9\left(\sqrt[3]{2}-1\right)}\approx1.709920934161365\dots,\label{eq:pw92-fzeta-exact}
\end{equation}
 \citeref{Perdew1992_PRB_13244} specifies the value $f''(0)=1.709\ 921$
for this quantity that is used in the interpolation, and this is the
value used in the Libxc implementation \texttt{lda\_c\_pw}, as well.
Employing the exact value of \ref{eq:pw92-fzeta-exact}, as in the
version called \texttt{lda\_c\_pw\_mod}, leads to slightly different
total energies, as is visible from \ref{tab:pw92-c}.

\begin{table}
\begin{tabular}{lrr}Functional & $ E_c(\text{N})/E_h $ & $ E_c(\text{Ne})/E_h $\\
\hline
\hline
\texttt{lda\_c\_pw} & $ -0.4563991 $ & $ -0.4947105 $\\
\texttt{lda\_c\_pw\_mod} & $ -0.4563981 $ & $ -0.4947091 $\\
\end{tabular}
\caption{Correlation energies of the N and Ne atoms for variants of the PW92 correlation functional, employing tabulated Hartree--Fock wave functions\cite{{Koga1999_IJQC_491}} evaluated with AtomicOrbitals\cite{{Furness__}} and a 2000 point radial quadrature with the default scheme of \citeref{Lehtola2022_JCP_174114}. Libxc keywords are used to identify the functionals.}
\label{tab:pw92-c}
\end{table}

Unfortunately, as the PW92 functional is an ingredient in many GGAs
and meta-GGAs, the choices for the employed value of $f''(0)$ can
also affect many other functionals. These include the PW91,\citep{Perdew1992_PRB_6671,Perdew1993_PRB_4978}
PBE,\citep{Perdew1996_PRL_3865,Perdew1997_PRL_1396} the B97 class
of functionals,\citep{Becke1997_JCP_8554} AM05,\citep{Armiento2005_PRB_85108,Mattsson2008_JCP_84714}
BMK,\citep{Boese2004_JCP_3405} GAPC,\citep{Fabiano2014_JCTC_2016}
and SOGGA11\citep{Peverati2011_JPCL_1991} GGAs, as well as the BC95\citep{Becke1996_CJC_995},
CC,\citep{Schmidt2014_JCP_18} M05,\citep{Zhao2005_JCP_161103} DLDF,\citep{Pernal2009_PRL_263201}
M08-HX and M08-SO,\citep{Zhao2008_JCTC_1849} M11,\citep{Peverati2011_JPCL_2810}
M11-L,\citep{Peverati2012_JPCL_117} MN12-L,\citep{Peverati2012_PCCP_13171}
MN12-SX,\citep{Peverati2012_PCCP_16187} MN-15,\citep{Yu2016_CS_5032}
MN15-L,\citep{Yu2016_JCTC_1280} revM11,\citep{Verma2019_JPCA_2966}
VSXC,\citep{VanVoorhis1998_JCP_400} B98,\citep{Becke1998_JCP_2092}
and CC06\citep{Cancio2006_PRB_81202} meta-GGA functionals. Functionals
that build on top of these functionals may also be affected; for instance,
the PKZB,\citep{Perdew1999_PRL_2544} TPSS,\citep{Tao2003_PRL_146401,Perdew2004_JCP_6898}
and SCAN\citep{Sun2015_PRL_36402} meta-GGAs all build on top of the
PBE expressions and as a result, one needs to be aware of the underlying
choice in each case. Many functionals assume the more precise value
of \ref{eq:pw92-fzeta-exact}, as it arises directly from \ref{eq:pw92-fzeta},
as is also the case with the $\omega$B97M-V functional,\citep{Mardirossian2016_JCP_214110}
for example.

A recent example of issues with the definition of $f''(0)$ is the
Google Advanced Science 2022 (GAS22) density functional,\citep{Ma2022_SA_279}
which can be described as a rediscovery of $\omega$B97M-V. The Jupyter
notebook cited in the supporting information of \citeref{Ma2022_SA_279}
contained a total energy for the Si atom in the def2-QZVPPD basis
set\citep{Weigend2005_PCCP_305}, which we wanted to reproduce with
the Libxc implementation.

After ensuring that the Libxc implementation used the more precise
parameters of the reference implementation,\cite{GAS22bug} and that
our calculations were converged to the quadrature grid limit, a significant
difference of several $\mu E_{h}$ still remained in the total energy
of the Si atom.

Examining the reference Jupyter notebook implementation revealed that
the truncated value for $f''(0)$ was used instead of the exact value
of \ref{eq:pw92-fzeta-exact} originally employed in $\omega$B97M-V,
as well as in our reimplementation of GAS22 in Libxc. In the def2-SVP\citep{Weigend2005_PCCP_305}
basis set, this inconsistency lead to a $3.1\ \mu E_{h}$ difference
in the total energy of the Si atom in the self-consistent calculations
with the reference implementation employing PySCF\citep{Sun2020_JCP_24109}
and our calculations with Libxc\citep{Lehtola2018_S_1} using ERKALE\citep{Lehtola2012_JCC_1572}.
When the same parameters were used, the difference between total self-consistent
energies produced by the two implementations was reduced to $5.5\ $n$E_{h}$.
\bibnote{See \url{https://gitlab.com/libxc/libxc/-/issues/419} (accessed 11 July 2023).} 

\paragraph{B3LYP}

There are several more examples of disparate functional forms in the
literature. We will comment on the perhaps most infamous one: the
B3LYP functional.\citep{Stephens1994_JPC_11623} This functional is
based on Becke's three-parameter hybrid functional\citep{Becke1993_JCP_5648}
(B3PW91)
\begin{equation}
E_{\text{xc}}^{\text{B3PW91}}=E_{\text{xc}}^{\text{LDA}}+a_{0}\Delta E_{x}^{\text{HF}}+a_{x}\Delta E_{x}^{\text{B88}}+a_{c}\Delta E_{c}^{\text{PW91}}\label{eq:b3pw91}
\end{equation}
where $\Delta E^{\text{HF}}=E_{x}^{\text{HF}}-E_{x}^{\text{LDA}}$
is an exact exchange correction, while $\Delta E_{x}^{\text{B88}}$
and $\Delta E_{c}^{\text{PW91}}$ are gradient corrections for B88
exchange and PW91 correlation. 

In B3LYP, \citet{Stephens1994_JPC_11623} replaced PW91 correlation,
which was not yet available in their program, with a combination of
the LYP\citep{Lee1988_PRB_785} GGA correlation functional and the
LDA correlation functional of \citet{Vosko1980_CJP_1200} (VWN).\bibnotemark[FrischCCL] \bibnotetext[FrischCCL]{Michael Frisch's email reply to Mikael Johansson's question on the Computational Chemistry List, see \url{http://www.ccl.net/chemistry/resources/messages/2002/05/22.008-dir/}. Accessed 26 April 2022.} 

Unfortunately, the paper by VWN describes more than one functional;
Libxc implements six variants that reproduce different energies, as
demonstrated by the data in \ref{tab:vwn-c}. Instead of the recommended
version, VWN5, which had been known in the literature as VWN for 14
years by the time B3LYP was published, the variant of VWN implemented
in the \texttt{\Gaussian{}} program is the RPA version, VWN(RPA),
which was also used in the B3LYP functional.

\begin{table}
\begin{tabular}{lrr}Functional & $ E_c(\text{N})/E_h $ & $ E_c(\text{Ne})/E_h $\\
\hline
\hline
\texttt{lda\_c\_vwn} & $ -0.4590053 $ & $ -0.4962448 $\\
\texttt{lda\_c\_vwn\_rpa} & $ -0.6593495 $ & $ -0.6670629 $\\
\texttt{lda\_c\_vwn\_1} & $ -0.4480821 $ & $ -0.4962448 $\\
\texttt{lda\_c\_vwn\_2} & $ -0.4579805 $ & $ -0.4962448 $\\
\texttt{lda\_c\_vwn\_3} & $ -0.4584408 $ & $ -0.4962448 $\\
\texttt{lda\_c\_vwn\_4} & $ -0.4611213 $ & $ -0.4962448 $\\
\end{tabular}
\caption{Correlation energies of the N and Ne atoms for variants of the VWN correlation functional, employing tabulated Hartree--Fock wave functions\cite{{Koga1999_IJQC_491}} evaluated with AtomicOrbitals\cite{{Furness__}} and a 2000 point radial quadrature with the default scheme of \citeref{Lehtola2022_JCP_174114}. Libxc keywords are used to identify the functionals.}
\label{tab:vwn-c}
\end{table}

This discrepancy has been the source of much confusion. Because VWN5
was the recommended variant in the literature, many other programs
implemented B3LYP with VWN5 instead of VWN(RPA)---the exact flavor
of the functional not having been specified in \citeref{Stephens1994_JPC_11623}---and
disagreements between the two were later found in the literature,
as has been discussed by \citet{Hertwig1997_CPL_345}, for example. 

We note the caveat that to this day, B3LYP is not the same functional
in all programs, and it is the user's responsibility to find out which
flavor is used. Both forms of the B3LYP functional are available in
Libxc: \texttt{hyb\_gga\_xc\_b3lyp} for the original version of \citet{Stephens1994_JPC_11623},
and \texttt{hyb\_gga\_xc\_b3lyp5} for the VWN5 variant. 

\paragraph{Becke's hybrids}

The half-and-half hybrid functionals originally introduced by Becke
are another example. The BHLYP functional was originally implemented
in \texttt{\Gaussian{}} incorrectly as half LDA plus half Hartree--Fock
exchange in combination with the LYP correlation functional (available
in Libxc as \texttt{hyb\_gga\_xc\_b}handh), while the correct composition
contains Becke'88 exchange\citep{Becke1988_PRA_3098} instead of LDA
exchange\citep{Bloch1929_ZfuP_545,Dirac1930_MPCPS_376} (\texttt{hyb\_gga\_xc\_bhandhlyp}).\citep{King1996_JPC_6061,King1996_JCP_6880,King1997_JCP_8536}
In some codes, such as Turbomole,\citep{Balasubramani2020_JCP_184107}
BHLYP refers to the latter form, which other codes offer as BHandHLYP.
As shown by the data in \ref{tab:becke-hyb}, these forms are clearly
not equal, and it is again the user's responsibility to know which
version is employed. Please note that the data in \ref{tab:becke-hyb}
only contains the DFA energy, thus excluding the exact exchange energy
from the Hartree--Fock component, which is the same in both functionals.

\begin{table}
\begin{tabular}{lrr}Functional & $ E_{xc}(\text{N})/E_h $ & $ E_{xc}(\text{Ne})/E_h $\\
\hline
\hline
\texttt{hyb\_gga\_xc\_bhandh} & $ -1.8141705 $ & $ -2.0647437 $\\
\texttt{hyb\_gga\_xc\_bhandhlyp} & $ -2.1502112 $ & $ -2.4160210 $\\
\end{tabular}
\caption{The DFA part of the exchange-correlation energies of the N and Ne atoms for the BHLYP and BHandHLYP functionals, employing tabulated Hartree--Fock wave functions\cite{{Koga1999_IJQC_491}} evaluated with AtomicOrbitals\cite{{Furness__}} and a 2000 point radial quadrature with the default scheme of \citeref{Lehtola2022_JCP_174114}. Libxc keywords are used to identify the functionals.}
\label{tab:becke-hyb}
\end{table}

\subsection{Self-consistent calculations \label{subsec:SCF-calculations}}

We have previously examined the convergence of radial quadratures
of the exchange-correlation energy for various functionals in \citeref{Lehtola2022_JCP_174114},
where we showed many recent DFAs to be ill-behaved. In this section,
we study SCF calculations with select density functionals, which are
used to demonstrate the need to converge the radial quadrature to
obtain reliable reference energies. As in \citeref{Lehtola2022_JCP_174114},
we will consider the Li, N, Ne, Na, P, and Ar atoms.

The calculations in this section employ Psi4 version 1.8,\citep{Smith2020_JCP_184108}
which by default uses the M4 radial grid of \citet{Treutler1995_JCP_346}
with Gauss--Chebyshev quadrature of the second kind. This grid was
found to be one of the better performing alternatives in \citeref{Lehtola2022_JCP_174114},
even though a modified version of the Gauss--Chebyshev quadrature
was employed in that work.

We choose PW92\citep{Bloch1929_ZfuP_545,Dirac1930_MPCPS_376,Perdew1992_PRB_13244}
(\texttt{lda\_x} with \texttt{lda\_c\_pw}) and PBE\citep{Perdew1996_PRL_3865,Perdew1997_PRL_1396}
(\texttt{gga\_x\_pbe} with \texttt{gga\_c\_pbe}) to represent LDA
and GGA functionals, which tend to converge rapidly to the grid limit,
and the TPSS\citep{Tao2003_PRL_146401,Perdew2004_JCP_6898} (\texttt{mgga\_x\_tpss}
with \texttt{mgga\_c\_tpss}), MS0\citep{Sun2012_JCP_51101} (\texttt{mgga\_x\_ms0}
with \texttt{gga\_c\_regtpss}), MVS\citep{Sun2015_PNASUSA_685} (\texttt{mgga\_x\_mvs}
with \texttt{gga\_c\_regtpss}), SCAN\citep{Sun2015_PRL_36402} (\texttt{mgga\_x\_scan}
with \texttt{mgga\_c\_scan}), and r$^{2}$SCAN\citep{Furness2020_JPCL_8208,Furness2020_JPCL_9248}
(\texttt{mgga\_x\_r2scan} with \texttt{mgga\_c\_r2scan}) to represent
meta-GGA functionals.

We start the density functional calculation from a preconverged HF
solution, as this allows a direct comparison to our earlier study:
examining the quadrature grid convergence of the density functional
total energy of the first SCF iteration is analogous to the studying
the grid convergence of in \citeref{Lehtola2022_JCP_174114}. The
only difference to \citeref{Lehtola2022_JCP_174114} is that the HF
wave function is determined on-the-fly in the Gaussian basis set employed
in this work, instead of employing pretabulated Hartree--Fock wave
functions in Slater-type orbital basis sets as in our previous work.
The pretabulated densities used in \citeref{Lehtola2022_JCP_174114}
were discussed above in \ref{subsec:Tabulated-wave-functions}, and
we also used them in \ref{subsec:Studies-at-fixed} to demonstrate
the importance of using a consistent set of parameters in density
functionals.

To enable a tight convergence assessment similar to \citeref{Lehtola2022_JCP_174114},
tight convergence thresholds (\texttt{e\_convergence 1e-10} and \texttt{d\_convergence
1e-9}) were used for both the HF and DFT calculations. Our calculations
employ an angular Lebedev grid\citep{Lebedev1976_UCMMP_10} of 434
points, as calculations employing the 590-point grid yielded similar
results; after all, the atoms considered here were chosen due to their
spherically symmetric electron densities.

In addition, we turned off grid weight screening in the Psi4 calculations
(\texttt{dft\_weights\_tolerance -1.0}), as by default points with
small quadrature weights are thrown out. We examined the importance
of the basis function screening threshold on the grid, but the default
value of this threshold (\texttt{dft\_basis\_tolerance 1e-12}) appeared
to yield converged results.

The comparison of the grid convergence of the total energies $E$
\begin{equation}
\Delta E(N_{\text{rad}})=E(N_{\text{rad}})-E(1500)\label{eq:deltaE}
\end{equation}
evaluated with the HF density of the first iteration or the SCF density
for the functional of the final iteration reveals that the behavior
is similar in both cases. Significant differences in the behavior
are only observed when the quadrature approaches machine precision,
at which point the use of SCF orbitals introduces a small amount of
further numerical noise.

The convergence was also seen to be similar across basis sets: we
examined the split-valence polarized def2-SVP\citep{Weigend2005_PCCP_305}
and the triple-$\zeta$ polarized def2-TZVP\citep{Weigend2005_PCCP_305}
basis sets, as well as the extended, benchmark-quality AHGBS-9 basis
set.\citep{Lehtola2020_JCP_134108} Results for all calculations are
available in the Supporting Information; we only present our main
findings here, which we choose to exemplify with SCF calculations
in the def2-SVP basis set.

\begin{figure*}
\subfloat[PBE]{\begin{centering}
\includegraphics[width=0.5\textwidth]{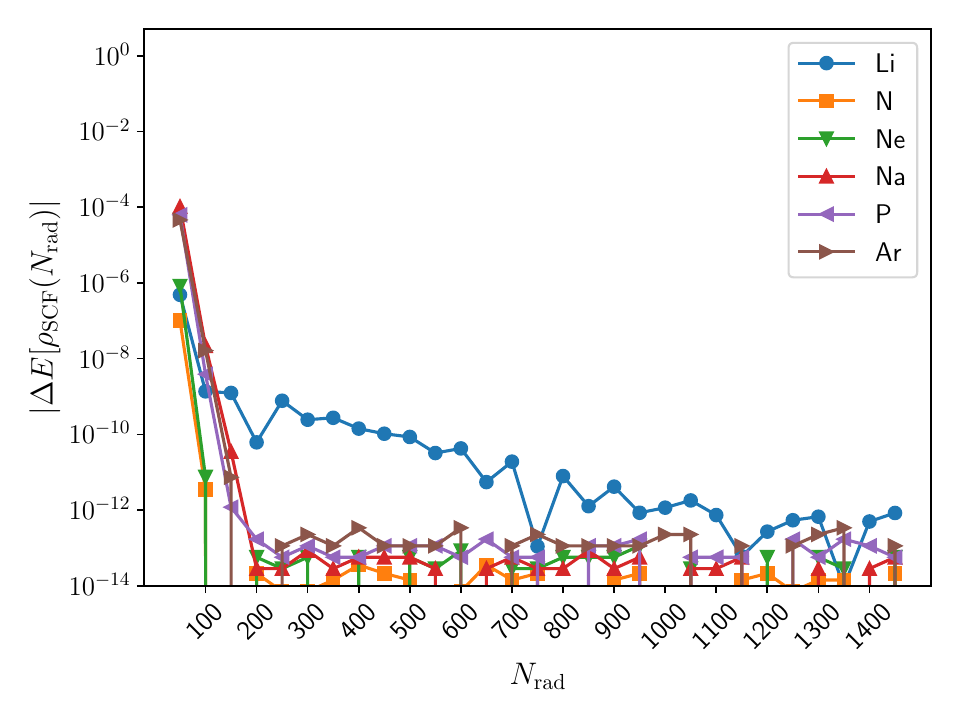}
\par\end{centering}
}\subfloat[TPSS]{\begin{centering}
\includegraphics[width=0.5\textwidth]{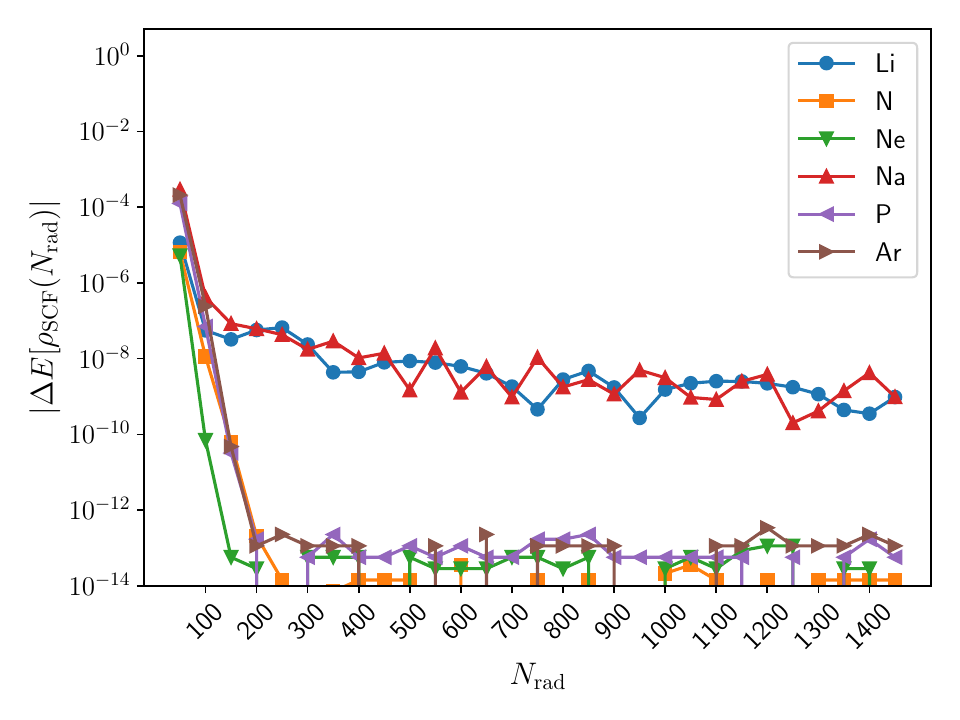}
\par\end{centering}
}

\subfloat[MS0]{\begin{centering}
\includegraphics[width=0.5\textwidth]{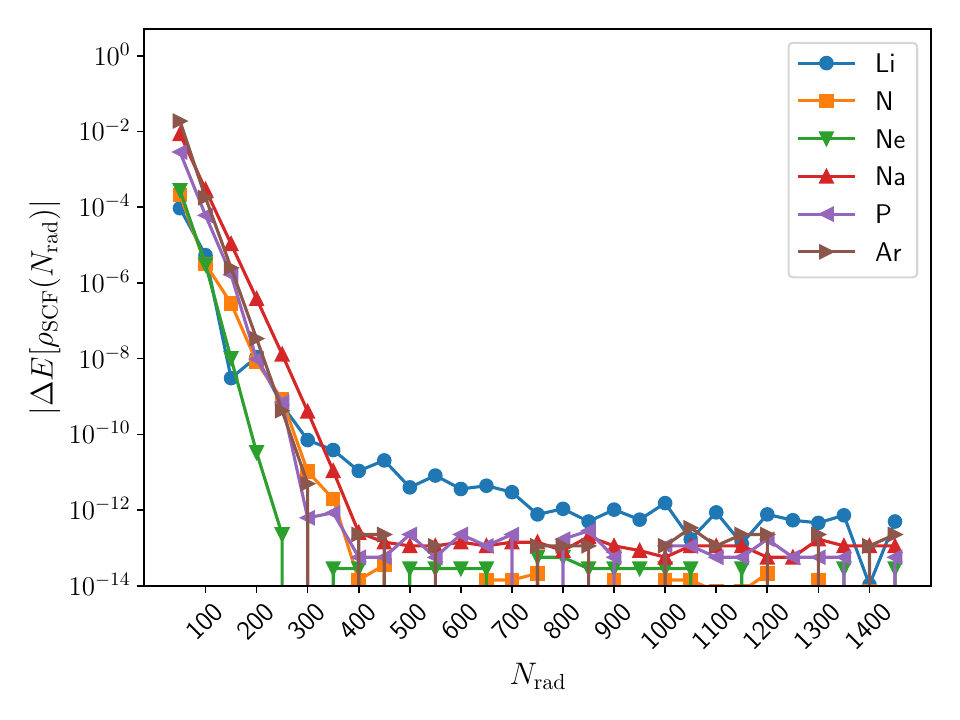}
\par\end{centering}
}\subfloat[MVS]{\begin{centering}
\includegraphics[width=0.5\textwidth]{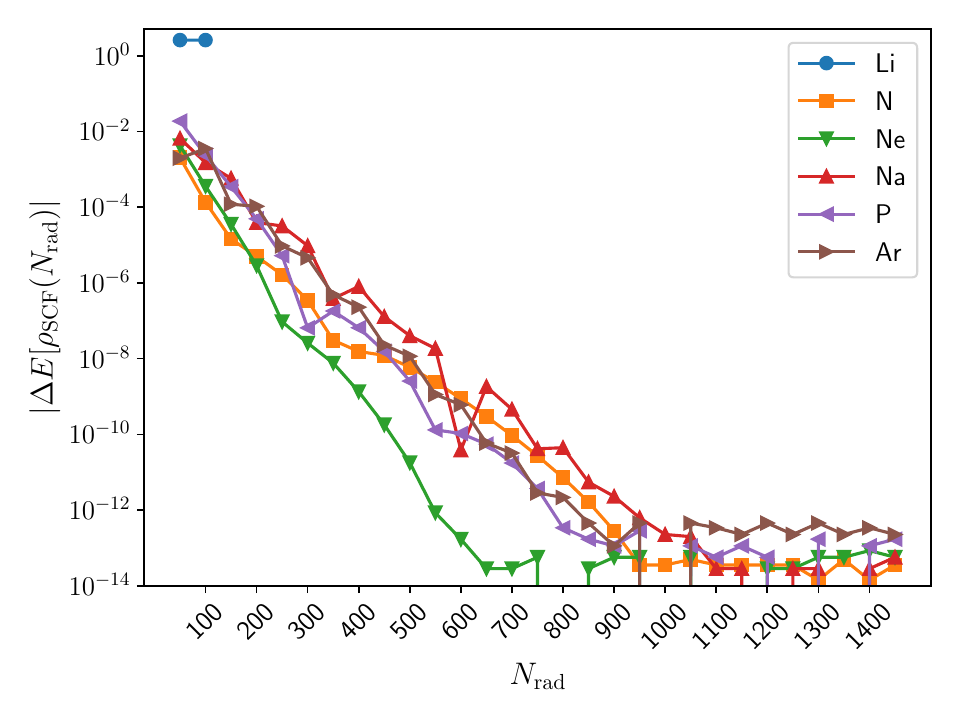}
\par\end{centering}
}

\subfloat[SCAN]{\begin{centering}
\includegraphics[width=0.5\textwidth]{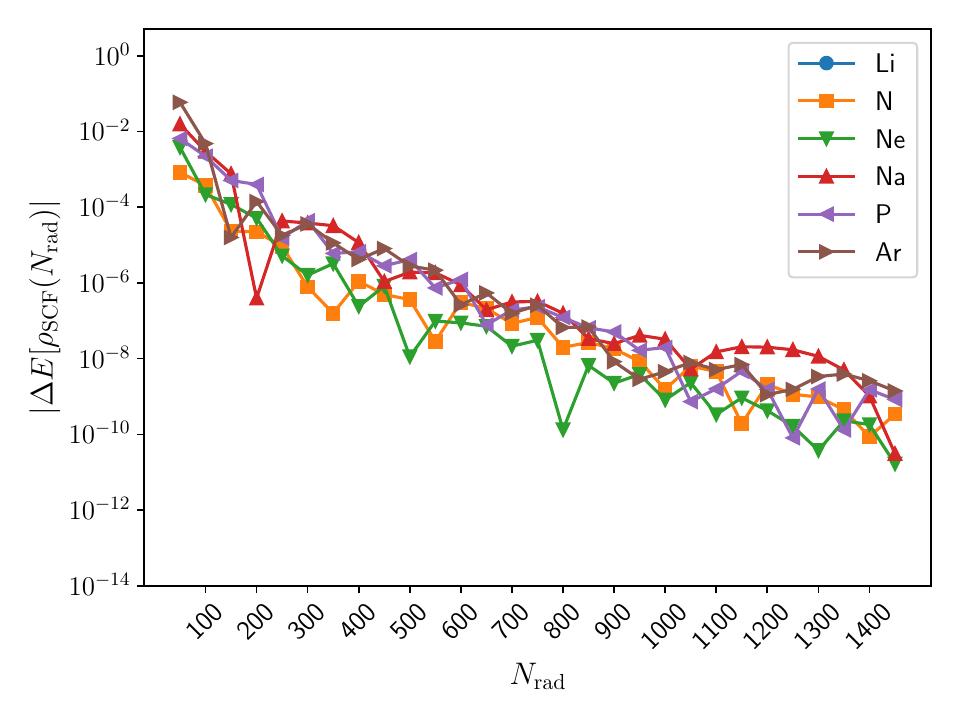}
\par\end{centering}
}\subfloat[r$^{2}$SCAN]{\begin{centering}
\includegraphics[width=0.5\textwidth]{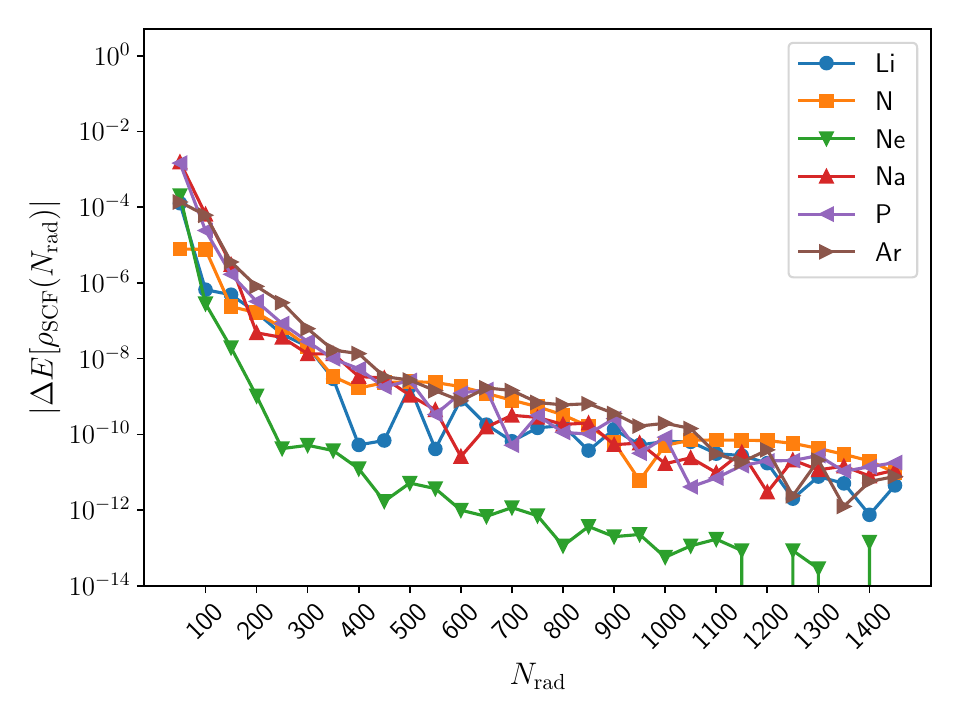}
\par\end{centering}
}

\caption{Grid convergence of the total energy determined with \ref{eq:deltaE},
evaluated with the SCF electron density in the def2-SVP basis set.
All points represent final total energies from separate SCF calculations.
\label{fig:SCF-def2SVP}}
\end{figure*}

The PW92 data turned out to be uninteresting due to rapid convergence
to machine precision. The data for the remaining functionals are shown
in \ref{fig:SCF-def2SVP}. As the total energy is seen to converge
in a similar manner in the present self-consistent calculations as
in the fixed-density evaluations of our previous work,\citep{Lehtola2022_JCP_174114}
these results emphatically confirm our analysis in \citeref{Lehtola2022_JCP_174114}:
the numerical well-behavedness of density functionals has not been
given sufficient attention in the past.

Although well-behaved functionals like PBE and TPSS (except for alkali
atoms) only require around 100 radial quadrature points to yield total
energies converged to $\mu E_{h}$ precision, others require hundreds
more. In the infamous case of SCAN, getting two programs to agree
on the energy to microhartree requires the use of 600--700 radial
quadrature points for the studied atoms... and this is assuming the
calculations converge, which may not be the case; the missing data
points for Li for MVS and SCAN in \ref{fig:SCF-def2SVP} are due to
lack of SCF convergence even in this small Gaussian basis set. These
two functionals are known to be unusably ill-behaved for fully numerical
calculations.\citep{Bartok2019_JCP_161101,Lehtola2023_JCTC_2502} 

We end this section with the note that the default radial grid in
Psi4 consists of 75 points, and that many other programs similarly
employ a default grid around this size. This again underlines that
a proper convergence study is required to determine reference energies
to sub-$\mu E_{h}$ precision.

\section{Summary and Discussion \label{sec:Summary-and-Discussion}}

As we have explained in this work, the history of density functional
approximations (DFAs) is full of independent implementations which
employ slightly different parameter values. In many cases, the origin
of the slightly different parameters is the unclarity of the original
literature, which allows several reasonable choices. 

The issue with such ambiguities is that they prevent the reproducibility
of the DFA: the total energy computed for a given DFA is not necessarily
directly comparable between different programs. Employing tabulated
Hartree--Fock electron densities for atoms, we have exemplified that
small changes to the numerical parameters employed in density functionals---representing
different choices for the parameters enabled by such ambiguities---can
have effects on total energies that are significant in applications
demanding high precision, such as fully numerical electronic structure
calculations. We underline that such different choices have been made
in the various implementations of many DFAs in different programs,
and that the reusable density functionals offered by Libxc are invaluable
in allowing the exact same DFA to be used across programs.

We have also extended our previous analysis of \citeref{Lehtola2022_JCP_174114}
with self-consistent calculations for the Li, N, Ne, Na, P, and Ar
atoms in various Gaussian basis sets. These calculations were used
to exemplify the need to converge the radial quadrature when reporting
reference energies for novel density functionals: as we have experienced
time and again, many functionals have been reported with unconverged
calculations. The self-consistent calculations also confirmed the
conclusions we made based on fixed densities in \citeref{Lehtola2022_JCP_174114}
that the numerical well-behavedness of new DFAs needs to be examined.

These two presentations combined with our analysis of various issues
in reported functionals in \ref{sec:problems-with-verification} point
out the need to furnish works publishing novel density functionals
with accurate reference energies that allow the verification of reimplementations
of the reported DFA. We have suggested several straightforward ways
in which to determine such data with publicly available free and open
source software.\citep{Lehtola2022_WIRCMS_1610} The systems for which
reference data is reported should include both spin-restricted and
spin-unrestricted systems; the N and Ne atoms offer excellent test
systems as they have well-behaved electronic structures.

Regardless of the employed approach, it is essential to converge the
calculation of the reference energy with respect to all numerical
parameters, most notably the quadrature grid, as we demonstrated in
\citeref{Lehtola2022_JCP_174114} and \ref{subsec:SCF-calculations}.
The reference energy should be computed and reported to very high
precision: using suitably large integration grids and small cutoff
thresholds, an agreement of better than 0.1 $\mu E_{h}$ in total
energies is typically achievable in Gaussian-basis calculations across
programs.

Such precise computation of the total energy is already a decent assessment
of the numerical behavior of the density functional: in general, the
more rapidly the quadrature converges, the better-behaved the functional
is. However, the issue of numerical well-behavedness has not been
given adequate attention by the functional developer community, as
many functionals do not afford such quick convergence with respect
to the quadrature.\citep{Lehtola2022_JCP_174114} 

In addition, we note that a good density functional should also afford
stable convergence to the complete basis set (CBS) limit.\citep{Lehtola2022_JCP_174114}
However, we again note that several recent functionals fail this criterion,
as well.\citep{Lehtola2023_JPCA_4180,Lehtola2023_JCTC_2502} In fact,
both types of checks are necessary: for example, while many Minnesota
functionals appeared well-behaved in the quadrature study of \citeref{Lehtola2022_JCP_174114},
they were later found to be ill-behaved in self-consistent calculations
in extended fully numerical basis sets.\citep{Lehtola2023_JCTC_2502,Lehtola2023_JPCA_4180}
Due to the increasing importance of fully numerical methods in electronic
structure theory, the numerical behavior of new density functionals
should be checked with fully numerical calculations, which are nowadays
routinely possible in the free and open source \textsc{HelFEM} program,\citep{Lehtola2019_IJQC_25945,Lehtola2019_IJQC_25944,Lehtola2020_PRA_12516,Lehtola2023_JPCA_4180,Lehtola2023_JCTC_2502}
for example. 

We end this work with a summary of our suggestions to ensure the reproducibility
of novel DFAs. We ask the following data to be furnished as part of
the publication of new density functionals:
\begin{enumerate}
\item the full mathematical equations for the functional, including all
the values for all the parameters with exactly the same values as
in the reference implementation
\item the source code for the reference implementation, if this implementation
is not already included in a standard open source library such as
Libxc or XCFun
\item reference energies for N and Ne atoms computed and reported to $0.1\mu E_{h}$
precision. 
\begin{enumerate}
\item For self-consistent calculations, the energies should be reported
separately for exchange-only calculations, and calculations with the
full exchange-correlation functional.
\item If the functional is defined by separate exchange and correlation
parts, total energies should be reported both for exchange-only calculations,
as well as calculations that include both exchange and correlation. 
\item For Gaussian-basis calculations, copies of the used Gaussian basis
set, supplemented with the Hartree--Fock energy computed in the basis
set.
\end{enumerate}
\end{enumerate}
We conclude with the statement that when novel functionals are originally
implemented as part of a common open source framework such as Libxc\citep{Lehtola2018_S_1}
or XCFun\citep{Ekstroem2010_JCTC_1971}, this greatly facilitates
the reproduction of results, because the same implementation is available
across a wide variety of programs. We invite the density functional
developer community to interact more strongly with standard libraries,
as including a new functional in such libraries makes it available
to a huge community of potential users.

\section*{Acknowledgments}

We thank Gustavo Scuseria, Viktor Staroverov and Giovanni Scalmani
(Gaussian Inc) for help in reproducing the PBE and TPSS functionals
in \Gaussian{}. We thank Cesar Proetto for pointing out an error
in one of the PW91 equations. We thank the National Science Foundation
for financial support under grant no. CHE-2136142. We thank the Academy
of Finland for financial support under project numbers 350282 and
353749.

\section*{Supporting Information}

Convergence plots of the total energies of the Li, N, Ne, Na, P, and
Ar atoms in the def2-SVP, def2-TZVP, and AHGBS-9 basis sets, employing
the HF and the SCF electron densities.

\bibliography{citations}

\begin{thebibliography}{155}
\bibitem [{1()}]{1}%
  \BibitemOpen
  \href@noop {} {}\bibinfo {note} {This error cancellation is also the reason
  why slightly different implementations of a DFA still tend to yield similar
  geometries and excitation energies, for example.}\BibitemShut {Stop}%
\end{thebibliography}

\end{document}